\documentclass[12pt]{iopart}
 \expandafter\let\csname equation*\endcsname\relax
 \expandafter\let\csname endequation*\endcsname\relax
\usepackage{amsmath} % Advanced math typesetting
\usepackage[utf8]{inputenc} % Unicode support (Umlauts etc.)
\usepackage[english]{babel} % Change hyphenation rules
\usepackage{array}
\usepackage{graphicx} % Add pictures to your document
\usepackage{listings} % Source code formatting and highlighting
\usepackage{color}
\usepackage{ulem}
\usepackage[titletoc]{appendix}
\usepackage{color}
\usepackage{hyperref}
\hypersetup{
    colorlinks=true,
    linkcolor=blue,
    urlcolor=red,
    linktoc=all
}

\begin{document}
\title{Dynamical and structural signatures of the glass transition in emulsions}
% \\ Experiments and Simulations}
\author{Chi Zhang $^{1,*}$, Nicoletta Gnan$^{2,3*}$, Thomas G. Mason $^{4,5}$, Emanuela Zaccarelli$^{2,3}$ and Frank Scheffold$^1$} 
\address{$^1$ Department of Physics, University of Fribourg}
\address{$^2$ CNR-ISC UOS  Sapienza, Piazzale A. Moro 2, 00185 Roma, Italy }
\address{$^3$ Department of Physics,  Sapienza University of Rome, Piazzale A. Moro 2, 00185 Roma, Italy }
\address{$^4$ Department of Chemistry and Biochemistry, University of California Los Angeles, Los Angeles,CA 90095, USA }
\address{$^5$ Department of Physics and Astronomy, University of California Los Angeles, Los Angeles, CA 90095, USA}
\address{$^*$ These authors contributed equally to this work}
\eads{\mailto{chi.zhang2@unifr.ch}, \mailto{nicoletta.gnan@roma1.infn.it}}

\begin{abstract}
We investigate structural and dynamical properties of moderately polydisperse emulsions across an extended range of droplet volume fractions $\phi$, encompassing fluid and glassy states up to jamming. Combining experiments and  simulations, we show that when $\phi$ approaches the glass transition volume fraction $\phi_g$, dynamical heterogeneities and amorphous order arise within the emulsion. In particular, we find an increasing number of clusters of particles having five-fold symmetry (\textit{i.e.} the so-called locally favoured structures, LFS) as $\phi$ approaches $\phi_g$, saturating to a roughly constant value in the glassy regime. However, contrary to previous studies, we do not observe a corresponding growth of medium-range crystalline order; instead, the emergence of LFS is decoupled from the appearance of more ordered regions in our system.  We also find that the static correlation lengths associated with the LFS and with the fastest particles can be successfully related to the relaxation time of the system. By contrast, this does not hold for the length associated with the orientational order. Our study reveals the existence of a link between dynamics and structure close to the glass transition even in the absence of crystalline precursors or crystallization. Furthermore, the quantitative agreement between our confocal microscopy experiments and Brownian dynamics simulations indicates that emulsions are and will continue to be important model systems for the investigation of the glass transition and beyond. 
\end{abstract}

\addtocontents{toc}{\protect\hypertarget{toc}{}}
\tableofcontents

\maketitle
Emulsions are of great practical importance in pharmaceutical, cosmetics, food, and agrochemical products \cite{petsev2004emulsions}.
In addition to their utility from an engineering perspective, these systems are gaining renewed attention in soft matter physics as model system for soft colloids \cite{vlassopoulos2014tunable,goyon2008spatial,jorjadze2013microscopic,scheffold2014jamming}.
Indeed, thanks to the ability of constituent droplets to deform without coalescing and thereby increase their surface area, emulsions can pack well beyond the so called random close packing or jamming limit \cite{torquato2000random,liu2010jamming,PhysRevLett.116.068302}, which represents the maximal volume fraction of hard-spheres when they are packed in a disordered manner. Crossing into and beyond random close packing, emulsions undergo a transition toward an elastic amorphous solid in which the rearrangement of droplets is not possible, the viscosity diverges and bulk samples exhibit a finite elastic shear modulus \cite{mewis2012colloidal,mason1995elasticity, mason1997osmotic,jop2012microscale,jorjadze2013microscopic,paredes2013rheology,scheffold2014jamming,scheffold2013linear}. Together with the huge effort in characterizing the jamming transition, great attention has been devoted to the lower density regime where emulsions behave similarly to a colloidal viscous fluid, displaying a  transition from an ergodic fluid state to a non-ergodic weak solid known as the glass transition \cite{crassous2006thermosensitive,scheffold2013linear,MasonJOSA1997,gang1999emulsion,goyon2008spatial,mason2014crossover,golde2016correlation}.  
\newline \indent As a general feature, when the glass transition is approached, the viscosity of a material sharply increases and the dynamics dramatically slows down. These phenomena are accompanied by the emergence of dynamical heterogeneities in 
the collective rearrangement of particles\cite{weeks2000three, berthier2011dynamical, ZhaoMasonPNAS}.
Despite the presence of heterogenous dynamics, structural quantities such as the radial distribution function $g(r)$ change very little suggesting the absence of large spatial correlations. 
The connection between structure and dynamics close to the glass transitions is a debated issue which has been discussed in different theroretical frameworks \cite{kirkpatrick1989scaling, biroli2008thermodynamic,toninelli2005dynamical,hedges2009dynamic}. More recently it has been shown that it is possible to link  the heterogeneous dynamics of particles with peculiar structural arrangements arising within the system on approaching the dynamic arrest \cite{kawasaki2007correlation,watanabe2008direct, tanaka2010critical,royall2015role,golde2016correlation}. The underlying idea is that regions of slow particle rearrangements must be connected to local structures which are energetically favorable. Such stable structures are thought to correspond to local minima of the energy landscape in which the system remains trapped. 
This is the picture proposed by Frank \cite{frank1952supercooling}, who has identified locally favourable structures (LFS) in the Lennard-Jones system as structures with icosahedral order. LFS maximise the density of packing and thus are energetically favoured with respect to the equilibrium face centred cubic (FCC) crystal structure of the Lennard-Jones system. In addition, geometric frustration introduced by their five-fold symmetry is incompatible with long-range order, hence the presence of LFS has been conjectured to have a fundamental role in the vitrification process \cite{tarjus2005frustration}.

Other numerical and experimental studies have revealed that crystalline order may play a key role in the vitrification process even if crystallisation is avoided. In this case there is an underlying reference crystalline state to which a specific bond orientational order (BOO) is associated. Simulations and experiments \cite{leocmach2013importance, leocmach2012roles} have shown that, although translational order is avoided suppressing crystal formation,  bond orientational ordering is still present and it actually grows upon cooling, extending up to medium range.  For the case of slightly polydisperse hard-spheres \cite{leocmach2012roles} it was shown that BOO is related to local dynamics, i.e. particles belonging to arrangements with high orientational order are less mobile. All these findings suggest that the dynamic slowing down and the increasing dynamical heterogeneities towards the glass transition may have some structural bases.

 Despite the growing interest in the packing of soft spheres over the last decade, emulsions have been largely underestimated as a quantitative model system.
Few experimental works on emulsions have focused on the properties in the glassy and jammed regime under shear flow \cite{mason1995elasticity, scheffold2013linear, mason1997osmotic,goyon2008spatial}, while the glassy behaviour of oil-in-water emulsions in the absence of flow has been studied only by dynamic light scattering \cite{gang1999emulsion}.
The great advantage of emulsions is that the droplet volume fraction is well defined since the liquid within the droplet is incompressible even when particle deform. 
This makes them suitable candidates for a better comparison with numerical and theoretical descriptions both with respect to hard spheres, for which the 
packing fraction definition is often problematic \cite{poon2012measuring,royall2013search}, and to soft particles such as microgels or star polymers that may deswell or interpenetrate \cite{C5SM03001C,mohanty2014effective,vlassopoulos2004colloidal,mattsson2009soft,gasser2014form,phdthesisScotti,scheffold2010brushlike}. Moreover solid friction and entanglements cannot play a role in emulsions, whereas they might in solid particulate or microgel dispersions.

In this work, we report an extensive study of the structural and dynamical properties of emulsions from states below the glass transition volume fraction up to jamming and we compare 3D fluorescent microscopy measurements with numerical simulations.
We identify multiple signatures of the glass transition, analyzing both dynamical and structural quantities, highlighting a connection between the dynamic slowing down and growing dynamical and structural correlation lengths. 
Such link has been revealed by looking at the increasing population of LFS as well as of clusters of fast particles
on approaching the glass transition. However, the relatively large polydispersity of our samples, beyond the known terminal polydispersity above which a single-phase crystallization can occur \cite{fasolo2003equilibrium,zaccarelli2009crystallization}, differentiate our system from previous studies, because of the absence of the growth of locally crystalline regions. Last but not least, we find a quantitative agreement with numerical simulations,  opening the pathway for future quantitative predictions for different soft repulsive particle systems over the entire range of concentrations from the fluid to the jammed phase.

\section{Experimental methods}
\subsection{Sample Preparation}
We prepare stable uniform oil-in-water emulsions as described in \cite{zhang2015structure}. We start with a $3:1$ mixture by weight of polydimethylsiloxane (PDMS; viscosity $15 \sim 45$ mPa.s, density 1.006 g/mL) and polyphenylmethylsiloxane (PPMS-AR200; viscosity 200 mPa.s, density 1.05 g/mL) and we emulsify it  in a couette shear-cell with sodium dodecyl sulfate
(SDS) surfactant in water for stabilizing the droplets. To remove evaporable
short molecules the PDMS oils is placed in an oven at $60^{\circ}C$ overnight prior to emulsification.
Depletion sedimentation \cite{bibette1991depletion} is then used to fractionate the droplets by size, until the desired polydispersity $PD\simeq12\%$ is achieved in the sample.
For such polydispersity  we find the size distribution of  droplets to be close to
log-normal with a mean droplet radius $a= 1.05$ $\mu$m or a droplet diameter $\sigma=2.1$ $\mu$m.
To sterically stabilize the droplets, SDS is replaced by the block-copolymer surfactant Pluronic F108. In addition,  formamide and
dimethylacetamid (DMAC) are added to the solvent in order
to simultaneously match the solute-solvent density and the refractive index at room temperature $T$ = 22 $^{\circ}$C.
Finally, the fluorescent dye Nile red is added to the solution in order to obtain optical contrast between the droplet and the dispersion medium.
Several hundred microliters of sample are spun down marginally above jamming  with centrifugation. The latter is carried out at 4 $^{\circ}$C in order to induce a slight density mismatch between the droplets and the solvent.
The stock sample then is diluted continuously in steps of $0.5\%$. After each dilution, we put a small amount
of suspension in an evaporation-proof  cylindrical cell of diameter $d=2$ mm and heigth $h=120$ $\mu$m sealed with UV-glue to a 
microscope cover slip.  
\subsection{Image acquisition}
$3$D High-resolution images  of droplets are obtained using a 
laser-scanning confocal microscopy  module, Nikon A1R, controlled by Nikon Elements software. Images are acquired with a X60 oil immersion objective with zoom X2.
Although the dye is present both in the continous phase and in the dispersed oil droplets, the emission spectra are different, exhibiting emission peaks of $670$ nm and $580$ nm
for the continuous phase and for the oil, respectively, when the sample is excited with a $488$ nm laser.
The dimension of the recorded images are $512$x$512$x$101$ pixels with a resolution of 0.21 $\mu m$/pixel in each direction.
Droplets are reconstructed by a template based particle tracking method
known as the Sphere Matching Method (SMM) \cite{brujic2004experimental} and Voronoi radical tessellation \cite{rycroft2009voro} to identify neighbors of each particle.
In our case the accuracy of the coordinates is roughly 20 nm in the lateral direction and 35 nm in the
vertical direction. The accuracy of the size determined from the analysis of immobile droplets is about 20 nm. Due to the finite exposure time, the size of droplets extracted from the tracking algorithm is slightly different for samples having different volume fractions, due to motion blurring of droplets in the images. Since all samples are made out of the same stock suspension, we assume that the particle size distribution for different samples is the
same. We proceed in the following way: first the particle size distribution of the system is measured at random close packing assuming $\phi_J$ = $64.2\%$, which corresponds to the value found for marginally jammed polydisperse frictionless spheres with polydispersity $PD\simeq12\%$ \cite{desmond2014influence,zhang2015structure}. This size distribution is  fixed throughout. Next, the particle size distributions obtained from the SMM tracking algorithm at lower concentrations are calibrated using this reference distribution, and, in this way, the volume fraction of each sample is determined. 
\newline \indent The fact that we can reversibly jam the system provides a well defined benchmark. In turn, we obtain a much better estimate for the absolute values of the packing fraction as compared to hard sphere systems \cite{poon2012measuring}. We estimate the absolute accuracy of our $\phi$-values to be better than $0.5\%$ with a statistical error better than $0.3\%$. The small difference is due to the finite systematic error with respect to $\phi_J$ \cite{desmond2014influence,zhang2015structure}. 

\section{Simulation methods}
\begin{figure}[!htbp]
\centering \includegraphics[width=0.7\linewidth]{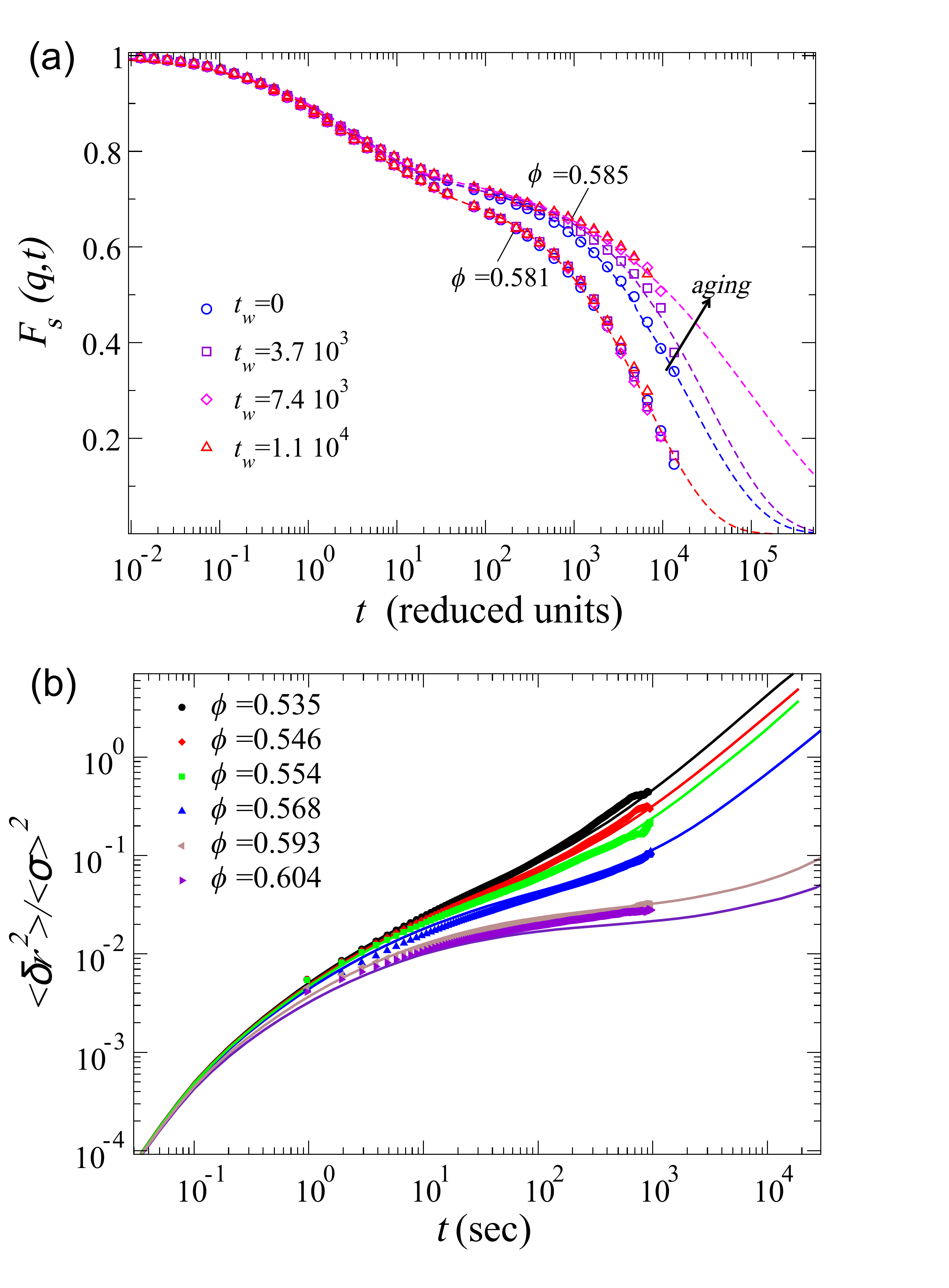}
\caption{(a) Self intermediate scattering function (ISF) for two droplet volume fractions $\phi=58.1\%$ and $\phi=58.5\%$ evaluated at different waiting times $t_w$ and $q<\sigma>\simeq 7.2$. While for $\phi=58.1\%$ non aging effects are observed, for $\phi=58.5\%$ an increase of the relaxation time of the self ISF as a function of $t_w$ is not negligible. Dashed lines are guides to the eye. (b) Normalized mean square displacements. Droplet volume fractions range from $53.5\%$ to $60.4\%$. Symbols are data from confocal microscopy measurements; lines are results from Brownian dynamics simulations. Note that the simulation curves have been shifted on the time-axis by the same arbitrary factor to match the experimental microscopic timescale.}
\label{fig:msd}
\end{figure}

To model the behaviour of dense emulsions we use a soft repulsive potential, following previous works\cite{lacasse1996model, berthier2009compressing, ikeda2012unified, ikeda2013disentangling, scheffold2013linear, vlassopoulos2014tunable} which have shown how the elastic and dynamic properties across the glass and the jamming transitions depend not only on the volume fraction $\phi$ but also on the strength of the repulsion. We thus model emulsions as particles interacting with a harmonic potential
\begin{equation}
\beta U(r_{ij})=\, u_0\,(1-r/\sigma_{ij})^2\Theta(r_{ij} -\sigma_{ij})
\label{eq:harmonic}
\end{equation}

\noindent where $i,j$  is the index of two particles with diameter $\sigma_i$ and $\sigma_j$  (with $\sigma_{ij}=0.5(\sigma_{i}+\sigma_{j})$) and $u_0$ is proportional to the harmonic spring constant and is in units of $k_BT$. The length unit is chosen to be the
average colloid diameter $\langle\sigma\rangle$ and time $t$ is in units of  $\langle\sigma\rangle \sqrt{m/u_0}$ (reduced units) where $m$ is the mass of a single particle. We perform Brownian Dynamics (BD) simulations of $N=2000$ polydisperse particles; a velocity Verlet integrator is used to integrate the equations of motion with a time step $dt=10^{-4}$. 
% A velocity Verlet integrator is used to integrate the equations of motion with a time step $dt=10^{-4}$.
We follow Ref. \cite{russo2009reversible}  to model Brownian diffusion by defining the probability $p$ that a particle undergoes a random collision every $X$ time-steps for each particle. By tuning $p$ it is possible to obtain the desired free particle diffusion coefficient $D_0=(k_BTX dt/m)(1/p -1/2)$.
We fix $D_0=0.0081$ in reduced units, for which the crossover from ballistic
to diffusive regime, for isolated particles, takes place at $t\sim 0.01$.
\newline \indent Using a harmonic approximation for the interaction potential is reasonable for small deformations of the droplets \cite{lacasse1996model,seth2006elastic}. Here, we consider only concentrations at or below jamming $\phi\le \phi_J=64.2\%$ \cite{zhang2015structure} and thus deformations can be considered small and the harmonic potential approximately applies. As discussed in a previous work \cite{scheffold2013linear}, the value of $u_0$ is set by the surface tension of the system. Indeed  a $1\%$ change in volume fraction above random close packing corresponds to a droplet compression $(1-r/\sigma)^2=2.7\cdot10^{-5}$, thus as long as  $u_0$ is much larger than $10^{5} k_BT$ the energy cost to thermally induce a corresponding shape fluctuation is $ \gg k_BT$. Based on these considerations, we set $u_0 =1.0\cdot 10^7$, which also matches rheology data Ref. \cite{scheffold2013linear,elasticityRheo}. For such values of $u_0$, the system under study is hard enough to be considered almost as hard-spheres, since the droplet deformation due to thermal fluctuations is very small. Nonetheless, the softness of the droplets and the absence of friction are key properties of emulsions that allow for the preparation of dense and marginally jammed systems. The polydispersity of the system is described by a log-normal distribution with unitary mean and standard deviation equal to $PD=12\%$ following the experimental probability size distribution.
The total simulation time for all the volume fractions investigated ranges  between $5.5 \cdot10^{7}$ and $2.4\cdot 10^{8}$ BD steps, corresponding to $t\in[5.5 \cdot10^{3}, 2.4\cdot 10^{4}$] in reduced units.
A recent numerical work on HS with polydispersity $\simeq12\%$ \cite{zaccarelli2015polydispersity} has shown that the relaxation features of the system depends very much on the population of small and large particles belonging to the tails of the size distribution. In the HS system, aging also affects fully decaying intermediate scattering functions (ISF) when $\phi>59\%$, which depend not only on the observation time $t$ but also on the waiting time $t_w$, i.e. the time elapsed from the beginning of the experiment or simulation.
Due to polydispersity, it was found that small and large particles undergo a dynamical arrest at different packing fractions; while large HS particles are dynamically arrested already at $\phi=58\%$, small particles are still free to move in the matrix formed by the large particles. For our emulsions, we observe a similar behavior, finding that the system starts to display aging for $\phi>58.1\%$. This is shown in Fig. \ref{fig:msd} (a) where the self ISF defined as $F_s(\vec{q},t)=(1/N)\sum_{1=1}^{N} e^{i\vec{q}\cdot(\vec{r}_i(0)-\vec{r}_i(t))}$ is displayed at different waiting times for $\phi=58.1\%$ and $\phi=58.5\%$ and wave vector $\vec{q}$ roughly corresponding to the position of the first peak of the structure factor $S(q)$. Hence, for $\phi>58.1\%$ we consider the system to be out-of-equilibrium.

\section{Comparison between numerical and experimental data}
Using Brownian dynamics (BD), rather than molecular dynamics (MD), in the simulation method is advantageous, because BD yields more accurate microscopic dynamics of emulsion droplets, thereby enabling us to achieve very good quantitative agreement between numerical and experimental dynamical observables, such as the mean square displacement, over an extended dynamic range in time. The use of MD simulations would have only allowed us to compare the resulting transport coefficients, such as the long-time diffusion coefficient $D$, although with better numerical efficiency in terms of computational time.  
By contrast to other systems \cite{zaccarelli2015polydispersity}, the determination of the packing fraction does not require any adjustable free parameter, and we directly use the experimental values in the simulations.
In order to improve the agreement reported with experimental data, we had to account in simulations for the error in the experimental exposure time, which is a source of noise in the coordinates along the three axis in confocal microscopy measurements. In fact, the scan over a single particle takes on average $1$ s, a time in which the particle is free to explore a certain volume within the cage.
 As a consequence, the  coordinates of particles extracted are affected by a noise that results in a suppression of the peak of the $g(r)$ \cite{mohanty2014effective}. 
 Since the short time motion for samples with different volume fraction is different, we would expect different level of noise on increasing $\phi$.
 An estimate of how much a particle with average radius $a\sim 1\mu m$ has moved in $1$s is given by the cage size which can be approximately written as \cite{doliwa1998cage,weeks2002properties}
 $\epsilon=4a[(\phi_J/\phi)^{1/3}-1]$. In addition to that, we consider the accuracy of the particle tracking. This brings an error of roughly $\delta_{track}\simeq 0.1$pixel (with $1$pixel $\simeq 0.21 \mu m$) in the lateral direction and $\delta_{track}\simeq 0.15$ pixel in the axial directions. Basing on such consideration, the noise can be approximately estimated as a Gaussian distribution $P(0, \epsilon^2 +\delta^2_{track})$ with zero mean and variance $w=\epsilon^2 +\delta^2_{track}$. We apply such Gaussian noise to  the  three coordinates of all the particles in simulations, finding a very good agreement with experimental results.
\section{Results}
\subsection{Dynamical properties close to the glass transition}
\subsubsection{Mean square displacement and $\alpha$-relaxation}
\begin{figure}[!htbp]
\centering \includegraphics[width=0.7\linewidth]{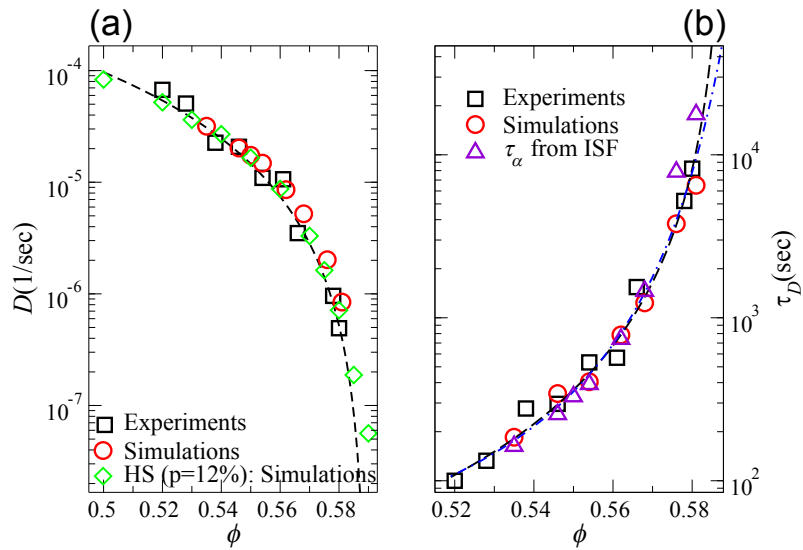}
\caption{(a) Diffusion coefficient $D$ as a function of the volume fraction $\phi$ for experiments (open squares) and simulations (open circles).  Open diamonds are numerical results for HS particles with polydispersity $PD=12\%$ from Ref.  \cite{zaccarelli2015polydispersity}. Note that numerical D values have been shifted by an arbitrary factor to match the experimental results for emulsions. The dashed line is the power-law fit of the experimental data set which gives $\phi_g=58.9\%$ and $\gamma=2.29$. The same interpolation for numerical data gives  $\phi_g=59.1\%$ and $\gamma=2.12$; (b) relaxation time $\tau_D$ extracted from the mean square displacement for both experiments (open squares) and simulations (open circles). A power-law fit of the two data sets gives, respectively, $\phi_g=58.9\%$ and $\gamma=2.1$ for experiments and $\phi_g=59.1\%$ and $\gamma=2.1$ for simulations. By interpolating experimental data with the VFT relation we obtain  $\phi_g=61.6\%$. The two interpolating lines for experimental data are shown in the figure (dashed lines). For comparison we also show the relaxation time $\tau_{\alpha}$ extracted from the numerical self ISF (open triangles). As in the left panel, numerical data have been shifted by an arbitrary factor. We find that $\tau_{\alpha}$ starts to decouple from the numerical $\tau_D$ on approaching the glass transition packing fraction.  }
\label{fig:relaxation}
\end{figure}

We start our discussion by showing the dynamical properties of emulsions in experiments and simulations around the glass transition volume fraction. Fig. \ref{fig:msd} (b) shows the comparison between the two sets of data for the mean square displacement  $\langle\delta r^2\rangle=(1/N)\sum_{i}|\vec{r}_i(0)-\vec{r}_i(t)|^2$. As for most of molecular liquids and colloidal systems, the dynamics shows a dramatic slowing down on approaching $\phi_g$ and $\langle\delta r^2\rangle$ displays the emergence of a typical plateau associated to the presence of "cages" in which particles remain trapped for an increasingly long time. We find that a simple model such as harmonic spheres quantitatively captures the dynamical behaviour of emulsions in an extended time region covering more than two decades for a wide range of packing fractions at and below jamming. Note that to superimpose experimental and numerical data a shift in time has been applied to the numerical mean square displacement. From the long-time limit of numerical mean-squared displacements we can extract the diffusion coefficient using the Einstein relation $D=\langle \delta r^2\rangle/6t$. For experiments, we could not reach a purely diffusive long-time regime, thus we estimate $D$ by introducing a relaxation time $\tau_{D}$, through the 
empirical relation $\langle \delta r^2\rangle =3(1+t/\tau_{D})\epsilon^2$\cite{ChiThesis} where $\epsilon$ is the characteristic cage size. The derivation of such expression is found in Appendix A. The associated diffusion coefficient is defined as $D=3\epsilon^2/\tau_D$.  The resulting numerical and experimental $D$ and $\tau_D$ are shown in  Fig. \ref{fig:relaxation}. 

\begin{figure}[!htbp]
\centering \includegraphics[width=0.7\linewidth]{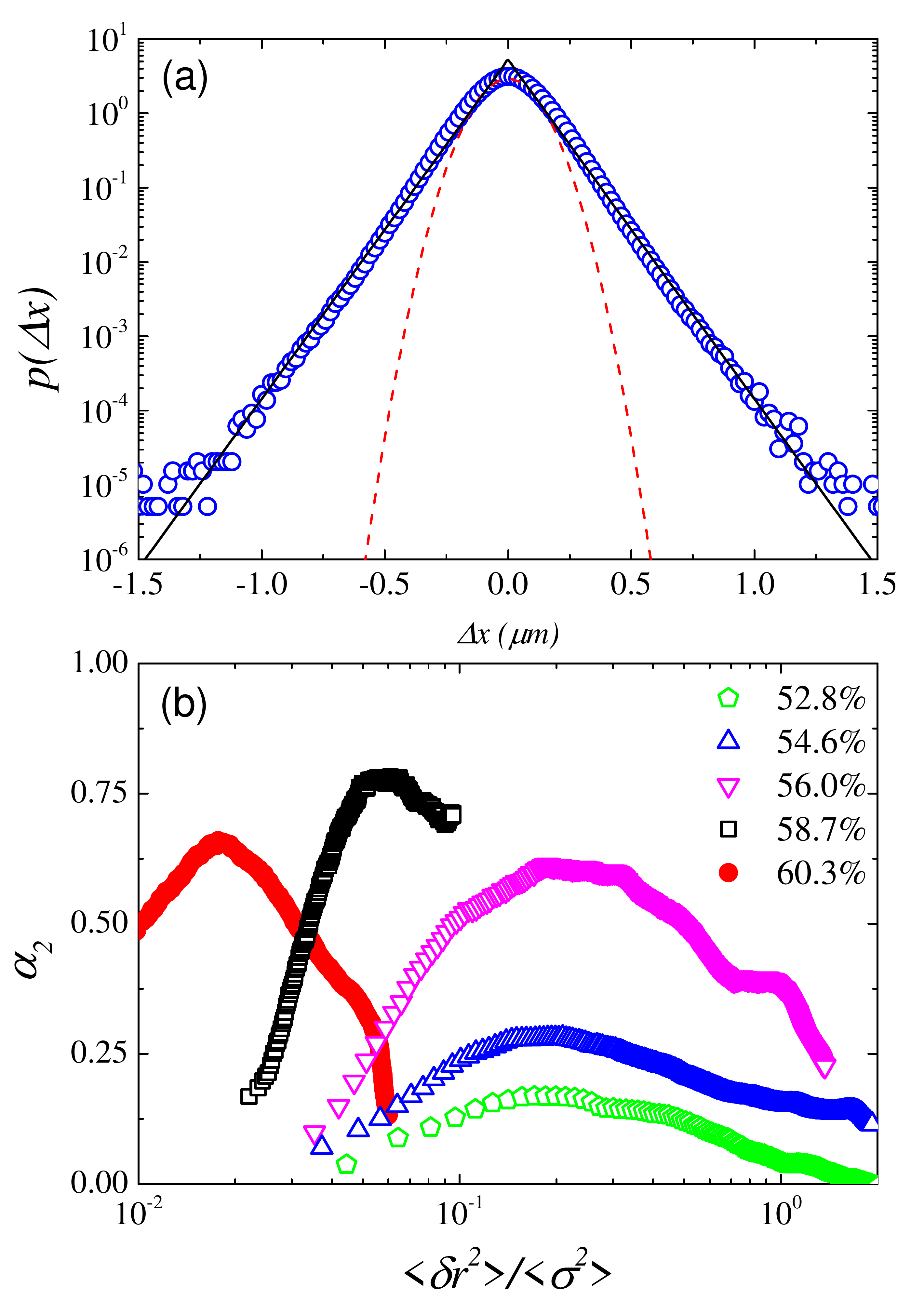}
\caption{Experimental measurements of non-Gaussian step-size distributions and dynamical heterogeneities for $\phi$ near and above $\phi_g$. (a) Distribution of droplet displacements $\Delta x$ obtained from confocal microscopy at a volume fraction of $\phi=60.3\%$ and time $t=2160$ s. Dashed line: best fit of the peak center to a Gaussian distribution; solid line: best fit of the tails to a stretched exponential distribution. (b) Non-Gaussian parameter $\alpha_2$ as a function of the dimensionless mean square displacement for volume fractions from $52.8\%$ to $60.3\%$.}
\label{fig:a2}
\end{figure}

The diffusion coefficient is represented in the Fig. \ref{fig:relaxation} (a), showing no difference on the way it has been calculated (Einstein or empirical relation). We also notice that
the  results are in good agreement with previous numerical data for  a HS sphere system with the same polydispersity \cite{zaccarelli2015polydispersity} that we plot together with results from emulsions, to show that our system behaves almost as HS.  By performing a power-law fit $D\propto |\phi-\phi_g|^{\gamma}$ we find that  $\phi_g=58.9\%$ and $\gamma=2.29$ for experiments, while $\phi_g=59.1\%$ and $\gamma=2.12$ for simulations, which are both in good agreement with power-law fits of Ref. \cite{zaccarelli2015polydispersity}. However, differently from HS simulations, we do not observe a deviation from a power-law decay in our numerical study; this is because Brownian dynamics is slower than molecular dynamics and does not allow to probe, within the same simulation time, the time scales that can be explored with MD. Hence, we are more far from $\phi_g$ than in Ref.  \cite{zaccarelli2015polydispersity}, to observe any deviation. The relaxation time $\tau_D$ is shown  in Fig. \ref{fig:relaxation} (b); a power-law fit of $\tau_D$ as a function of $\phi$, gives similar results for $\phi_g$ and $\gamma$.  A slightly higher value of $\phi_g\simeq61.6\%$ is obtained if data are instead interpolated with the empirical Vogel-Fulcher-Tammann (VFT) expression
\begin{equation}
\tau_{D} = \exp(A\phi_g/|\phi-\phi_g|)
\label{VFT}
\end{equation}
The small difference between the results of the two interpolations is again a  consequence of the fact that both numerical and experimental results are too far  from $\phi_g$ to observe a difference between the interpolating relations and discern which is the the best between the two. The two fits  (power law and exponential) for the experimental data set are shown also in the figure.
Finally we want to point out the difference between the numerical $\tau_D$ and the $\alpha$-relaxation time $\tau_{\alpha}$ extracted from the self ISF from simulations which are both shown in the same panel (dashed and dash-dotted lines); we find that the two times can be superimposed for a wide range of packing fractions, but start to show a decoupling on approaching the glass transition, a signature of the the violation of the Stokes-Einstein relation occurring between $D$ and $\tau_{\alpha}$ close to $\phi_g$.
\subsubsection{Dynamical heterogeneity}
One common way to characterize dynamic heterogeneities is to look for deviations of the particle displacements compared to free diffusion \cite{kob1997dynamical,weeks2000three}. For a random diffusion process the displacement distribution $P(\Delta x,t)$ at a given time $t$ is a Gaussian with zero mean and a variance equal to the mean squared displacement. Collective and correlated displacements lead to dynamic heterogeneities and deviations from the Gaussian distribution as shown in Fig. \ref{fig:a2}(a). Such deviations can be quantified by a non-Gaussian parameter defined as:
\begin{equation}
\alpha_{2} = \frac{3 \langle \delta r^{4}\rangle}{5 \langle \delta r^{2}\rangle^{2}}-1,
\label{Ngaussian}
\end{equation}

In Fig. \ref{fig:a2}(b) we plot $\alpha_2$ as a function of the particle mean squared displacement. Initially the values are nearly zero in the liquid but acquire appreciable values when approaching the glass transition volume fraction. In this regime $\alpha_{2}$ displays a pronounced peak. This is because the movement of particles results from the combination of the intra-cage and inter-cage dynamics. At short time scales, the displacement is mostly due to intra-cage dynamics and the distribution is nearly Gaussian $\alpha_{2}\sim 0$. Collective rearrangements are associated with cage breaking in the glass. Thus the peak in $\alpha_{2}$ is related to the size of the cage. As the cage size gets compressed the maximum of $\alpha_{2}$ is shifted towards smaller values of $\langle \delta r^{2}\rangle$. At long times, or large  values of $\langle \delta r^{2}\rangle$, the displacements are due to a sum of many random cage breaking processes and the distribution becomes Gaussian again.
\newline \indent Another interesting way to analyze the collective particle motion is to look specifically at a  of particles that differ from the Gaussian. Following \cite{weeks2000three} we define the population of fast particles as the $5\%$ most mobile particles within a certain time interval, calculated with respect to $t=0$. The ratio of $5\%$ is chosen based on the fact that  the percentage of particles whose displacement  deviates from a Gaussian distribution is roughly $5\%$ (Fig. \ref{fig:a2} (a)) \cite{kob1997dynamical,donati1999spatial}.

\begin{figure}[!htbp]
\centering \includegraphics[width=0.7\linewidth]{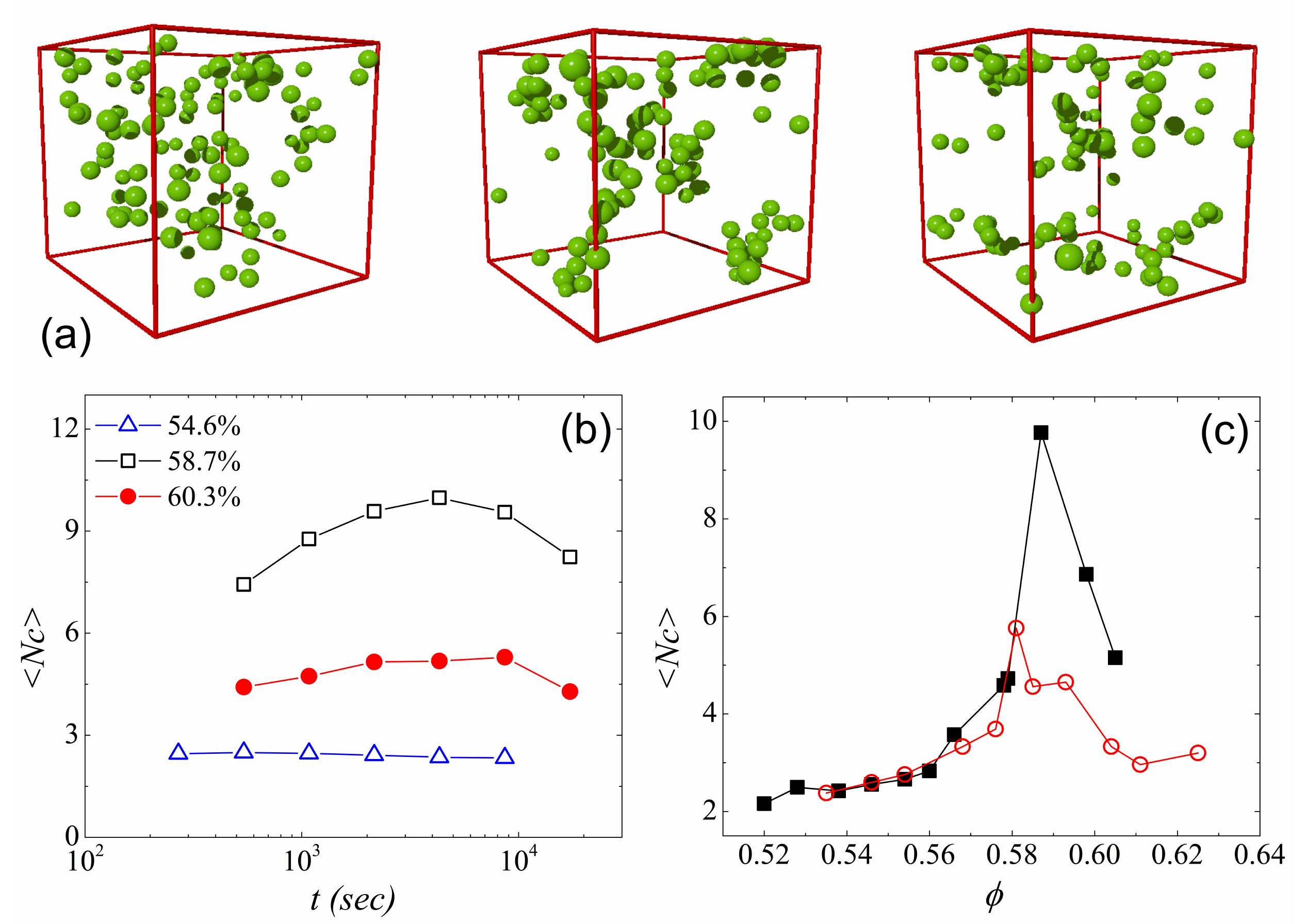}
\caption{(a) Identification of fast particles in the microscopy experiments at different volume fractions: from left to right $\phi=54.6\%$, $\phi=58.7\%$ and $\phi=60.4\%$ 
respectively for $t= 270 s, 4320 s, 8640s$. (b) Mean cluster size from experiments as a function of time for different samples with volume fraction of $54.6\%$, $58.7\%$ and $60.4\%$. Lines are guides to the eye. (c) Mean cluster size of fast particles as a function of the volume fraction $\phi$. Closed squares: experiments; open circles: simulations. } 
\label{fig:snapshot}
\end{figure}

Fig. \ref{fig:snapshot} shows several snapshots of fast particles that are spatially correlated. The appearance of spatial correlations is direct evidence for dynamic heterogeneities close to the glass transition \cite{weeks2000three,kegel2000direct}. We define clusters of $i$ particles from set of fast neighbouring particles identified via the Voronoi radical tesselation. The mean cluster size of fast particles is defined by taking the sum over clusters of all sizes and averaging over several configurations. The values for $\langle N_c\rangle$ we find depend both on concentration and on time . The latter is shown in Fig. \ref{fig:snapshot}. Especially close to $\phi_g$, $\langle N_c\rangle$ displays a pronounced maximum as a function of time. Moreover, this maximum is located close to the relaxation time $\tau_D$. Away from $\phi_g$ the peak is not pronounced or even absent. This suggests that, on approaching $\phi_g$, collective rearrangements play a increasingly important role.
By selecting the maximum value of $\langle N_c\rangle$ for several volume fractions, we can plot the concentration dependence of the cluster size $\langle N_c\rangle_{\textmd{max}}$ (In the absence of a clear maximum we select an arbitrary time). As shown in Fig. \ref{fig:snapshot} (b) and (c) the cluster size increases on approaching the glass transition and then decreases above $\phi_g$. This behaviour is observed both for simulations and experiments as shown in Fig. \ref{fig:snapshot}(a). We note that due to the polydispersity of the system small particles tend to be more mobile than larger particles. In connection to this it is worthwhile mentioning that the average size of the fast particle population is smaller, e.g. for $\phi=59.3\%$, the mean radius of fast particles is around 0.92$\mu$m, while for all particles it is 1.05$\mu$m. 
\subsection{Structural properties close to the glass transition}
\subsubsection{Radial distribution function}
\begin{figure}[!htbp]
\centering \includegraphics[width=0.8\linewidth]{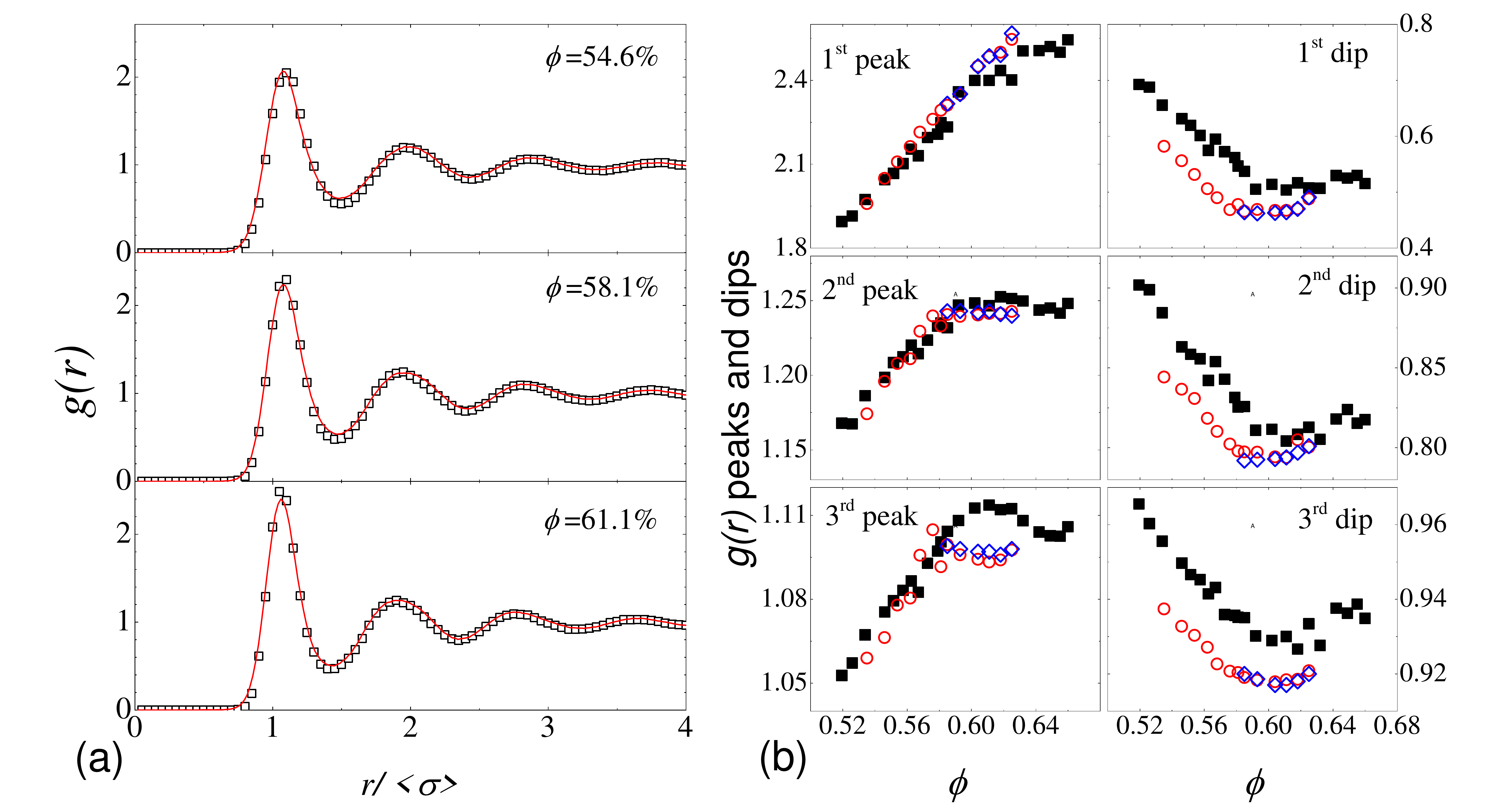}
\caption{(a) Radial distribution function $g(r)$ of soft spheres for three volume fractions $\phi$. Symbols denote experimental data for emulsions from confocal microscopy measurements, lines denote results from Brownian dynamics simulations. (b) Peak (left panel) and dip (right panel) amplitudes of $g(r)$ versus the volume fraction. Experiments: full squares. Simulations: open symbols. Circles show results obtained by averaging over a single run;  diamonds show results obtained by averaging over $100$ independent runs at time $t_w=2500$.}
\label{fig:grtotal}
\end{figure}

Fig. \ref{fig:grtotal} (a) shows the radial distribution functions of the system for three different packing fractions taken from experiments and simulations. The agreement is striking in the whole investigated range of packing fractions.

We thus analyze the concentration dependence of the minima and maxima of the $g(r)$ across the glass transition. The results are shown in Fig. \ref{fig:grtotal}  (b) for experiments and simulations. For the latter, the structural properties above $\phi_g$ have been obtained both by averaging over a single run (as for the experimental data) and by averaging over $100$ independent runs at a fixed waiting time $t_w=2500$, to eliminate the effects of aging in the sample. The two sets of data are displayed with different colours in Fig. \ref{fig:grtotal} (b), showing that the results are similar. 
The main interesting feature that we find is the non-monotonic behaviour of the peaks of the $g(r)$. 
While the first peak seems to be barely influenced by the presence of the glass transitions, the second and the third peak together with the first three dips of the $g(r)$ display a clear change at $\phi_g$. In fact we find that their amplitudes increase (peaks) or decrease (dips) on approaching the glass transition, saturating above $\phi_g$ meaning that the long-range structure remains unchanged by further compressing the emulsion. Differently, the behaviour of the first peak shows some changes within the first shell even beyond $\phi_g$ and seems to saturate only close to the jamming volume fraction \cite{zhang2015structure}. This is consistent with the behaviour found in other soft particles \cite{silbert2006structural, jacquin2011microscopic} such as PNIPAM particles \cite{zhang2009thermal} and granular materials \cite{cheng2010experimental}  close to jamming. In those cases a maximum in the first peak of the $g(r)$ has been predicted and experimentally observed as a structural signature of the jamming transition \cite{zhang2009thermal,paloli2013fluid,liu2010dynamical}. Our data is consistent with these previous studies. However, due to the onset of coalescence under significant droplet compression we cannot access deeply jammed samples ($\phi > 66\%$). The limited stability under compression is a trade-off when optimizing the emulsion systems for buoancy and index matching conditions. At and below  $\phi_J$ ($\sim 64.2\%$) we do not observe coalescence after more than one year.
\newline \indent The increase of the peaks and the decrease of all the dips of the $g(r)$ is related to the fact that, on increasing the volume fraction, particles tends to organize in better defined shells displaying a kind of  "amorphous order"\cite{biroli2008thermodynamic} that needs to be quantified. This picture can be captured by looking at those parameters that probe the local structure of the system. One parameter is the average number of neighbors.
\newline \indent There are different ways 
%how one can define
to determine the number of nearest neighbours. One possibility is to define a cut-off distance, such as the first minimum of the $g(r)$, $r_{min}$, and count all the neighbours within that distance from a specific particle $N_{coord}=4\pi \rho \int_0^{r_{min} } r^2 g(r)$. In that case the number of neighbours $N_{coord}$ is called coordination number. However, such a definition depends on the value of the cut-off that changes in dependence of the volume fraction. Here we implement a different approach. We consider two particles as neighbours if they share a wall of a Voronoi cell. This way, the result is unique and parameter free since it is based only on geometrical considerations. The Voronoi tessellation allows not only to count the number of neighbours but also to determine geometric properties of the cells as we will show later. The trends found in simulations and experiment are exactly the same and, except for a small shift, the data sets for $N$ superimpose as shown in Fig. \ref{fig:isoperimetric} (a).  
\newline \indent The concentration dependence of the average number of neighbours N shown in Fig. \ref{fig:isoperimetric} (a), reveals a clear change around $\phi_g$. When approaching $\phi_g$ the number of neighbours decreases. Above $\phi_g$ the average number of neighbours saturates close to the value predicted for random close packing $N=14.3$\cite{zhang2015structure}.
These observations can be rationalized by considering the evolution of the dips and peaks in the radial distribution function. Below $\phi_g$  the boundary between the first and the second neighbouring shell is shallow and the average number of neighbours found is thus larger. As the volume fraction increases, the two shells become well separated (the first dip of the $g(r)$ decreases) and, as a consequence, the average number of neighbours decreases. For  $\phi >\phi_g$ the first dip and all higher order dips and peaks saturate which is consistent with a constant number of neighbours in this regime. 
For comparison, the coordination number $N_{coord}$, extracted from the simulation data, is found to remain almost constant for $\phi <\phi_g$ and sharply decreases to a smaller value above the transition. 

\subsubsection{Isoperimetric quotient}
\begin{figure}[!htbp]
\centering \includegraphics[width=0.6\linewidth]{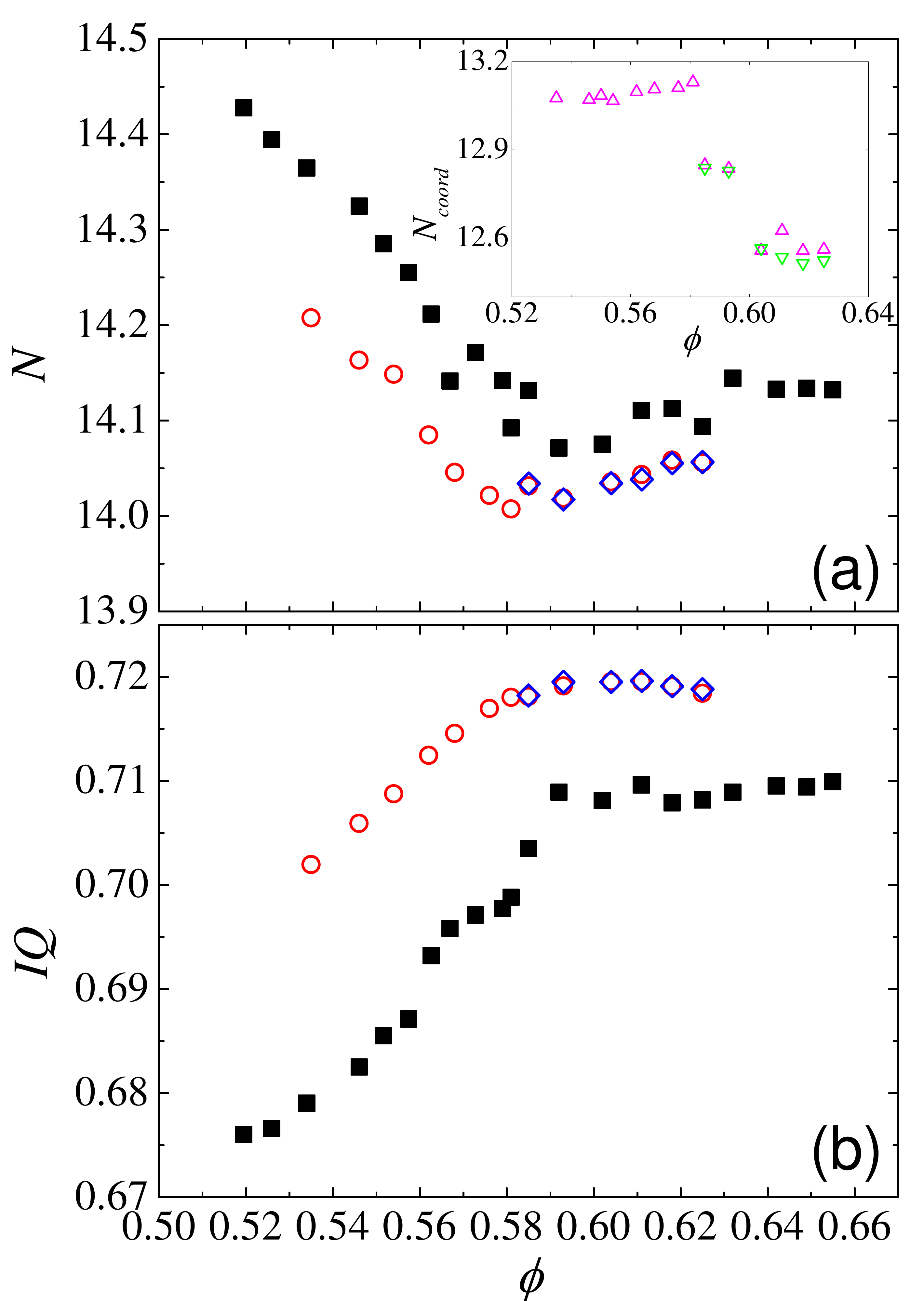}
\caption{Analysis of the Voronoi cell and the number of neighbors: (a) Average number of neighbors $N$ and (b) Isoperimetric quotient $IQ$ as a function of the droplet volume fraction $\phi$. Experiments: full squares. Simulations: open symbols. Circles are the result of a single run while diamonds are obtained by averaging over $100$ independent runs at $t_w=2500$. Inset in (a): Coordination number $N_{coord}$ from simulation data. Up triangles are the result of a single run while down triangles are obtained by averaging over $100$ independent runs at $t_w=2500$.}
\label{fig:isoperimetric}
\end{figure}
The isoperimetric quotient $IQ$ \cite{damasceno2012predictive} is an interesting measure that describes the similarity of a Voronoi cell to a sphere, and as such it is sensitive to shape changes of the cells. For an individual particle $i$, $IQ_i=36\pi V_i^2/S_i^3$ where $V_i$ and $S_i$ are the volume and the surface area of the Voronoi cell of particle i. $IQ_i$ is dependent on the configurations of the nearest neighbors, including the orientation and separation. With $IQ$ we denote the average of $IQ_i$ over all the particles. The evolution of $IQ$ as a function of the volume fraction $\phi$ is shown in Fig. \ref{fig:isoperimetric}(b). We find that the $IQ$ parameter increases up to $\phi_g$ indicating that the particles pack more homogeneously and thus tend to form more spherical Voronoi cells. Once the glass transition is approached, the packing geometry cannot be improved any further since an efficient particle rearrangement process is lacking. The saturation of $N$ and $IQ$ in the glass clearly shows that the geometrically configurations are frozen in and the only remaining process is the compression of the preformed cages until random close packing or jamming is reached at $\phi \to \phi_c$.

\subsubsection{Orientational correlation length}

\begin{figure}[!htbp]
\centering \includegraphics[width=0.6\linewidth]{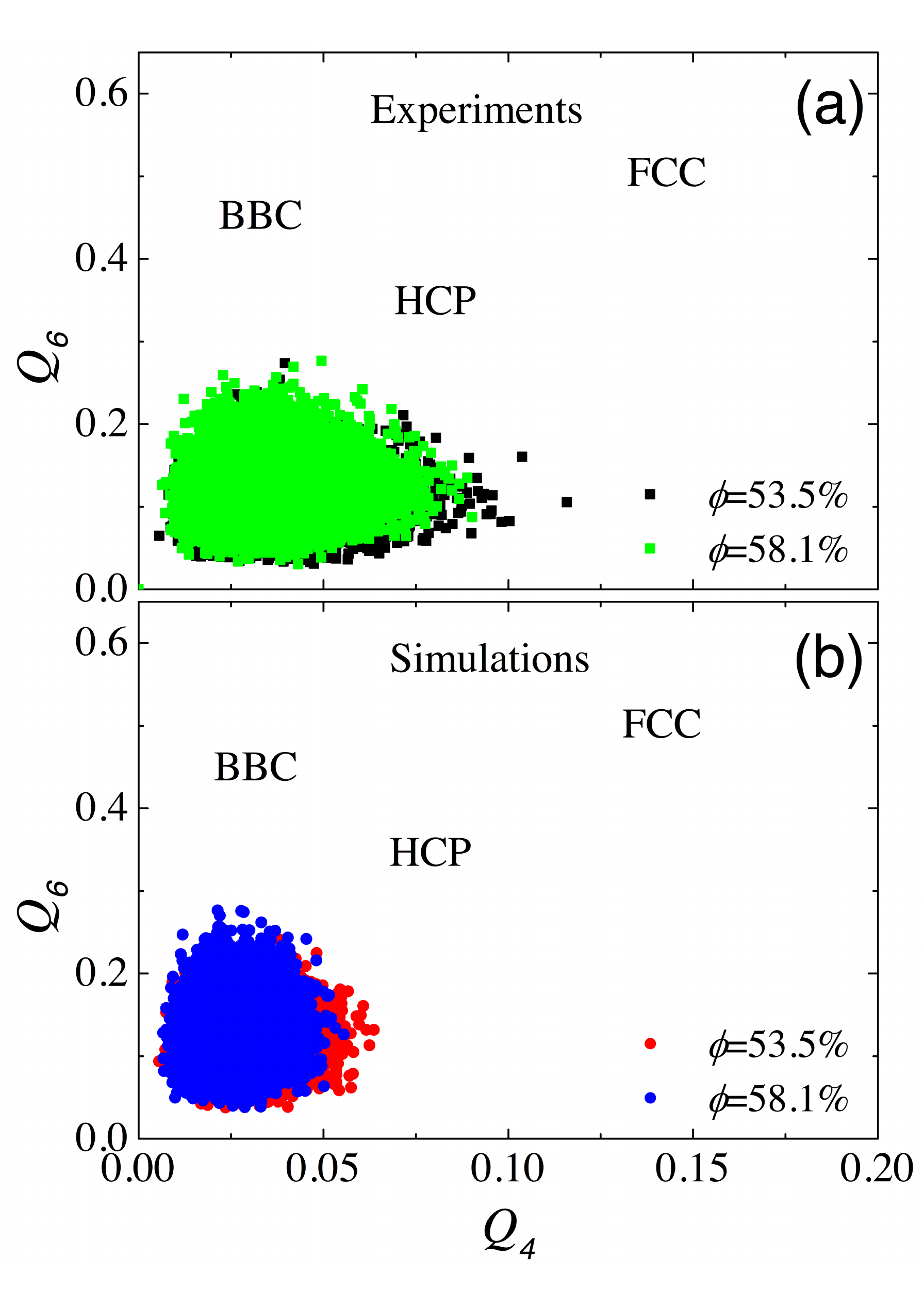}
\caption{Correlation map of bond orientational order (BOO) parameters $Q^{k}_4$
and $Q^{k}_6$ at two volume fractions. The figure highlights the regions in which  the order is related to three types of crystals commonly seen in colloidal system: FCC, BCC and HCP. (a) Experimental values for $\phi=53.5\%$ and $\phi=58.1\%$. (b) Simulations for $\phi=53.5\%$  and $\phi=58.1\%$ (obtained analysing $100$ independent configurations at $t_w=2500$). }
\label{fig:Q6vsQ4}
\end{figure}
Previous studies have suggested that dynamical heterogeneities are related to the emergence of a medium range crystalline order \cite{kawasaki2007correlation,tanaka2010critical,leocmach2012roles} in weakly polydisperse systems highlighted by a growing bond-orientational correlation length. Although such correlation has been found to grow in a "critical-like fashion", i.e. can be well fitted with some diverging law,  the correlation lengths observed are typically limited to few particle diameters only.
To investigate the presence of crystalline ordering in our moderately polydispersed emulsions, we use the bond orientational order parameters (BOO) which provide a powerful measure of the local and extended orientational symmetries in dense liquids and glasses \cite{steinhardt1983bond}. The BOO analysis focuses on bonds joining a particle and its neighbors. Bonds are defined as the lines that link together the centers of a particle and its nearest neighbors determined by Voronoi radical tessellation.
We define the BOO $l$-fold symmetry of a particle $k$ as the $2l+1$ vector:

\begin{equation}
q^{k}_{lm}=\frac{1}{N^k}\sum_{j=1}^{N^k} Y_{lm}(\Theta(\vec{r}_{kj}), \Phi(\vec{r}_{kj}))
\end{equation}

\noindent where $N^k$ is the number of bonds of particle $k$, $Y_{lm}(\Theta(\vec{r}_{kj}), \Phi(\vec{r}_{kj}))$ is the spherical harmonics of degree $l$ and order $m$ associated to each bond and $\Theta(\vec{r}_{kj})$ and $ \Phi(\vec{r}_{kj})$ are polar angles of the corresponding bond measured with respect to some reference direction. Following the work of Lechner and Dellago \cite{lechner2008accurate} we employ the BOO coarse-grained over the neighbours, which increases the accuracy of the type of medium-range crystalline order (e.g. FCC,HCP or BCC type):

\begin{equation}
Q^{k}_{l}=\sqrt{\frac{4\pi}{2l+1}\sum_{m=-l}^{l} |Q^{k}_{lm}|^2}
\end{equation}
\noindent with

\begin{equation}
Q^{k}_{lm}=\frac{1}{N^k_0}\sum_{j=1}^{N^k_0} q^{k}_{lm}(\vec{r}_{kj})
\end{equation}

\noindent and where $N^k_0$ is the number of nearest neighbors of particle k including particle k itself.
We first evaluate the behaviour of $Q^{k}_6$ and $Q^{k}_4$ which allow us to distinguish between cubic and hexagonal medium-range crystalline order. The results are shown for experiments and simulations respectively in Fig. \ref{fig:Q6vsQ4}(a) and (b). The correlation map of $Q^{k}_4$ and $Q^{k}_6$ reveals that, over the whole investigated range, only liquid-like structures are detected. This is due to polydispersity of our sample, which largely exceeds the known terminal polydispersity for single-phase crystallization in hard spheres \cite{fasolo2003equilibrium, zaccarelli2009crystallization, martinez2014exposing}. So far, only experimental results for weakly polydisperse hard spheres (with polydispersity around $6\%$) have been reported, in which a medium-range crystalline order of FCC type was observed. However in our system,the reference crystal phase is not trivial since particles should fractionate to crystallize \cite{fasolo2003equilibrium}. As a consequence also the BOO parameter does not reveal a clear tendency to organise in a specific crystal structure. Hence polydispersity in our case completely suppresses the formation of any crystal-like order even at the local scale.

\subsection{Locally favoured structures}
\begin{figure}[!htbp]
\centering \includegraphics[width=0.6\linewidth]{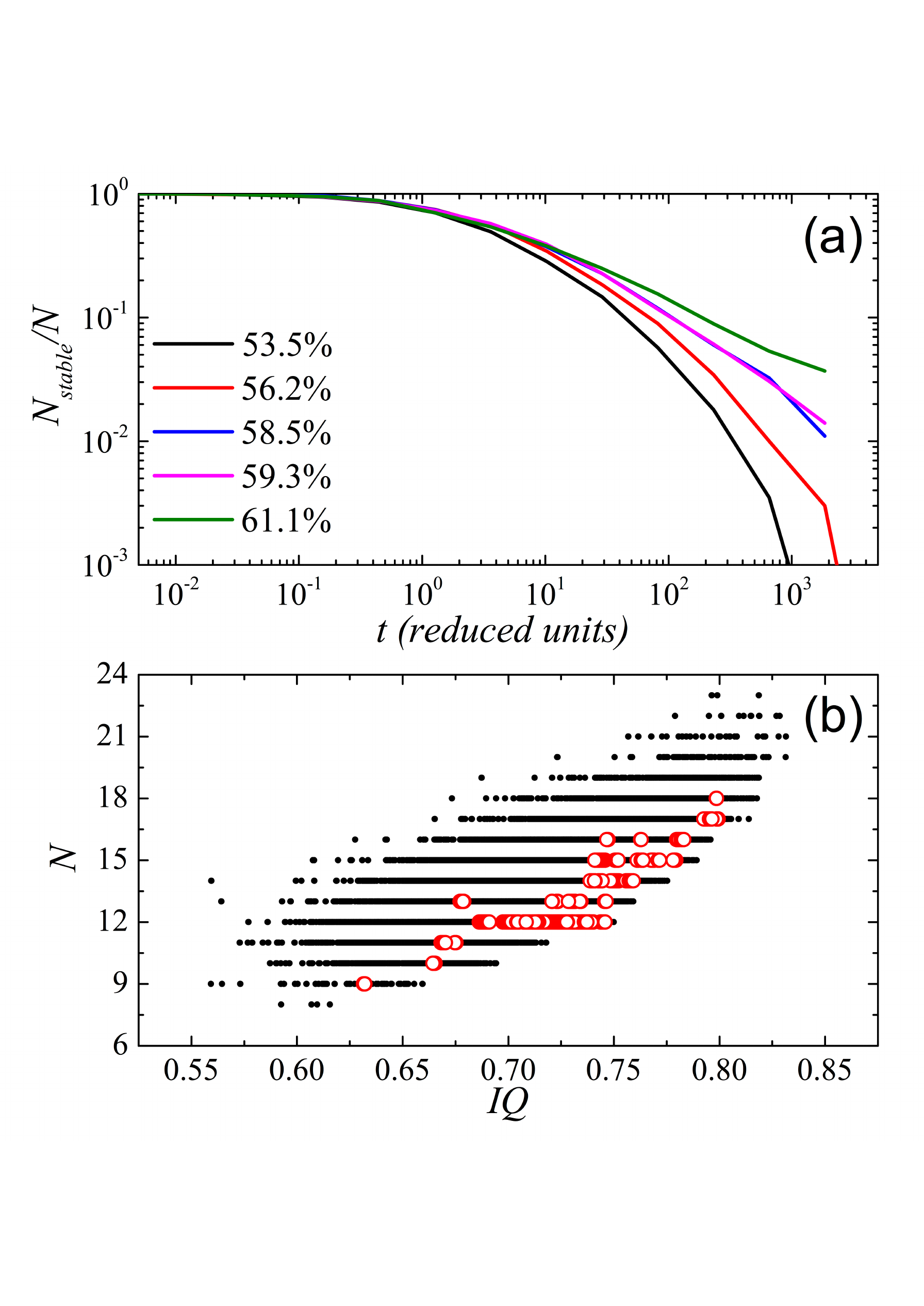}
\caption{ Decay of stable particle configurations for $\Delta t=2500$ in reduced units. (a) Simulation data showing the fraction of stable particles as a function of time for several volume fractions. (b) Map extracted from simulations for the number of neighbors $N$ versus the isoperimetric quotient $IQ$ for all particles (black dots) and for stable particles (open circles).}
\label{fig:structLFS-1}
\end{figure}
Locally favoured structures (LFS) are energetically favoured and as a consequence they should be longer-living in the system. LFS thus can be identified in the system by looking at the lifetime of the neighbours around a given particle. To this end we define a stable particle $i$ as the one that within a certain time interval $\Delta t$ maintains the same neighbours $n^{ij}$. The latter are defined 
as before via the Voronoi radical tessellation. In Fig. \ref{fig:structLFS-1}(a) we plot for different volume fractions the typical stable particle survival rate as a function of time defined as  $N_{stable}/N=\langle\sum_{i<j}n^{ij}(t)n^{ij}(t+\Delta t)\rangle/N(t)$, where $N$ is the total number of neighbours \cite{puertas2003simulation, zaccarelli2009colloidal}. We expect that, for a fixed $\Delta t$, on increasing $\phi$ the number of stable particles will increase since cage rearrangements become more difficult. For our analysis we fix $\Delta t=2500$ in reduced units.

\begin{figure}[!htbp]
\centering \includegraphics[width=0.6\linewidth]{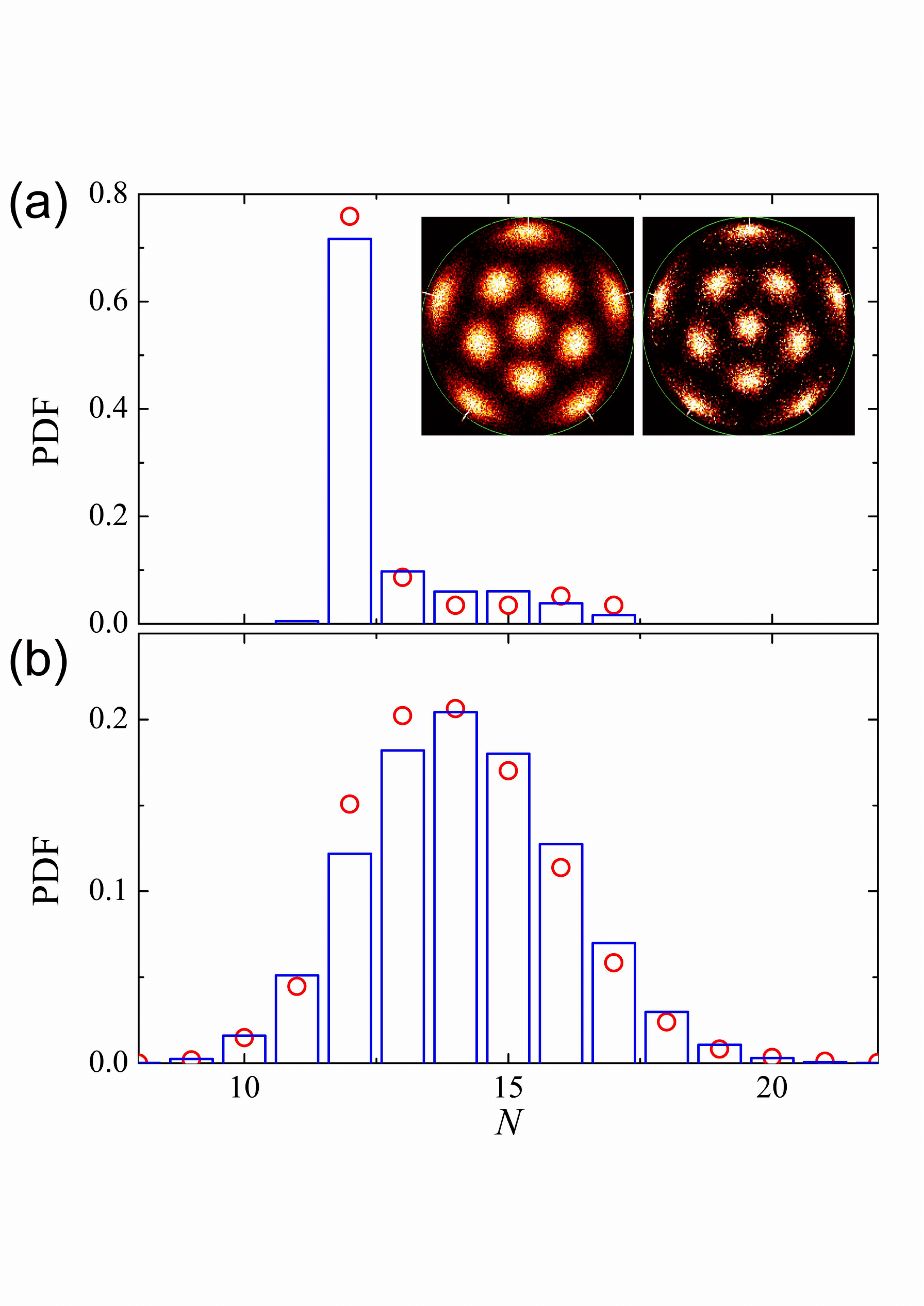}
\caption{ Distribution of number of neighbors for $\Delta t=2500$ in reduced units. (a) Stable particles, data averaged over $54.6\% <\phi < 60.4\%$. Experiments: bars; simulations: circles. (b) All particles at $\phi=58.1\%$. Experiments: blue bars; simulations: red circles. Inset: Bond-order diagram of $N$ = 12 particle clusters identified, as described in the text and Fig. \ref{fig:structLFS-2}. Left: experiments. Right: simulations.}
\label{fig:structLFS-2}
\end{figure}

\begin{figure}[h!]
\centering \includegraphics[width=0.6\linewidth]{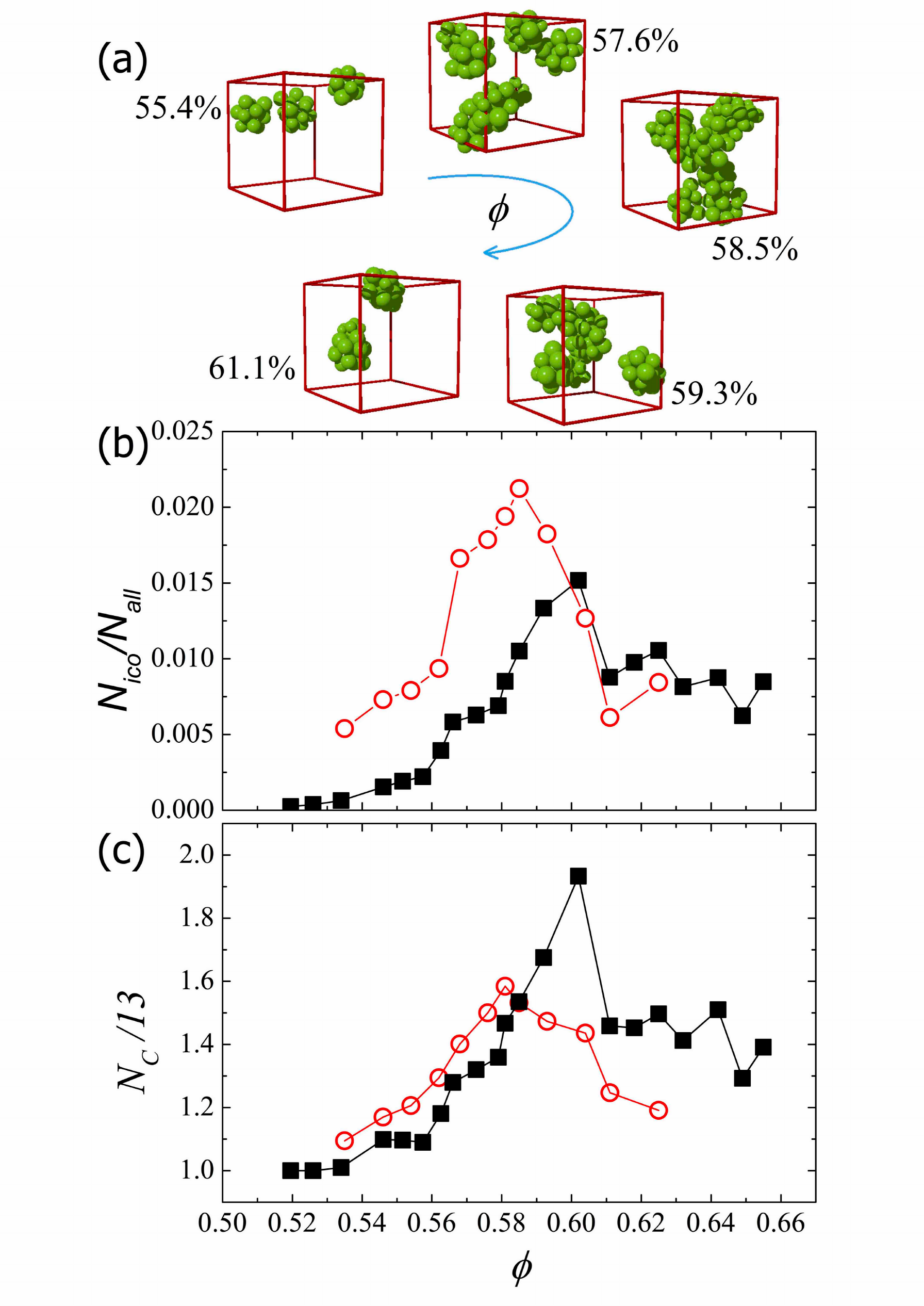}
\caption{(a) Visualization of typical LFS for different volume fractions (simulations). (b) Number of icosahedral centers over total number of particles. (c) Mean cluster size of icosahedral structures. Experiments: closed squares; simulations: open circles. Solid lines are guide to the eye. }
\label{fig:spatialLFS}
\end{figure}

\indent We first notice by looking at Fig. \ref{fig:structLFS-1}(b) that the number of neighbours of stable particles is strongly correlated with a high values of the isoperimetric quotient IQ : this suggests that stable particles belong to a peculiar structure with a given symmetry. This is confirmed in Figs. \ref{fig:structLFS-2} where the distribution of the number of neighbours is shown both for stable particles and for all particles. The comparison between the two distributions shows that most stable particles have exactly 12 neighbours both in simulations and in experiments, which is consistent with the idea that stable particles may form icosahedral structures. To verify this hypothesis we select stable particles with 12 neighbours and we perform a topological cluster classification (TCC)\cite{watanabe2008direct, malins2013identification} that allows to identify clusters that are topologically equivalent to certain reference clusters. The inset in Fig. \ref{fig:structLFS-2} (a) shows the bond-order diagrams of the $N=12$ particle clusters both for experiments and simulations. The ``heat style'' patterns stands for the probabilities of finding a neighbor in that direction. We start by considering that we are looking from the top of a icosahedral-structure with the central particle in the center of the figure.  The central spot shows the probability of finding the top neighbor. The first five-folded spots show the probability of finding the upper layer of five neighbors. The second five-folded spots shows the lower layer. The bottom neighbor is not shown.  Typical spatial icosahedral configurations of $N=12$ clusters for different volume fractions are shown in Fig. \ref{fig:spatialLFS}(a). 
 \indent The population of such structures is increasing when approaching some critical volume fraction around $\phi\sim 60\%$ as shown in Fig. \ref{fig:spatialLFS}(b). Here we plot the  fraction of the population of icosahedral centers as the volume fraction crosses $\phi_g$. In parallel the average size of connected clusters formed by icosahedral structures accordingly  increases as shown in Fig. \ref{fig:spatialLFS} (c). Here the cluster size $N_c$ is defined by considering
all particles which are part of icosahedral structures (both centers and neighbors) and thus for an unconnected, isolated cluster $N_c/13=1$. Therefore $Nc/13$ shown in Fig. \ref{fig:spatialLFS}(c) describes the cluster size normalized by a single icosahedral structure. Above the glass transition, both the overall number and the size of icosahedral domains decreases again.

\begin{figure}[h!]
\centering \includegraphics[width=0.6\linewidth]{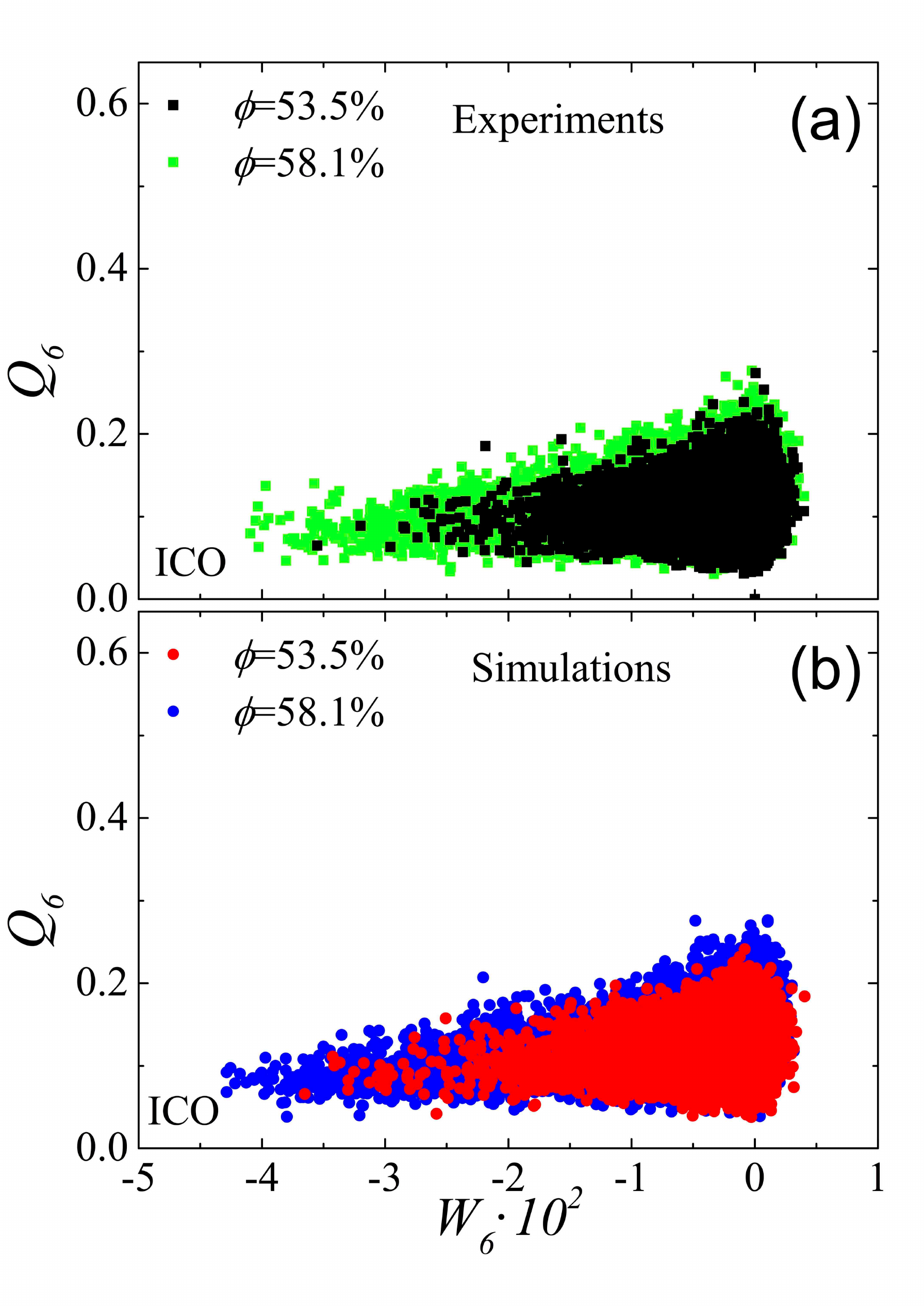}
\caption{Correlation map of bond orientational order parameters $w_6$
and $Q^{k}_6$ at two volume fractions. The figure highlights the regions in which  the order is related to icosahedral structures. (a) Experimental values for $\phi=53.5\%$ and $\phi=58.1\%$. (b) Simulations for $\phi=53.5\%$  and $\phi=58.1\%$ (obtained analysing $100$ independent configurations at $t_{w}=2500$).}
\label{fig:w6}
\end{figure}
\indent Finally in Fig. \ref{fig:w6} we consider correlations between the  BOO parameter Q6 and another order parameter called $w_6$, which is defined as

\begin{equation}
w_{6}(i)=\frac{\sum_{m_1+m_2+m_3}  \begin{bmatrix}
    6 & 6 & 6  \\
    m_1 & m_2 & m_3
  \end{bmatrix} q_{6m_1}(i) q_{6m_2}(i) q_{6m_3}(i) }{(\sum_{m=-6}^{6} |q_{6 m}(i)|^2)^{3/2}}.
\end{equation}
An increase of $w_6$ has been observed in polydisperse HS particles \cite{leocmach2012roles} together with the increase of crystalline order identified by a growth of the parameter $Q_6$. In our case, we do not observe an increase of $Q_6$, due to the higher polydispersity. Hence, contrary to what found in previous works on hard-spheres \cite{leocmach2012roles}, we observe that crystalline order remains modest while  icosahedral order grows when approaching the glass transition. We now need to understand if such growth is somehow related to the dynamic slowing down of the system close to $\phi_g$
%Hence, in our system, the onset of icosahedral structures can be decoupled from the appearance of crystal-like order.
\subsection{Link between structure and dynamics}
In the previous sections, we presented evidence of both dynamical and structural signatures
of the glass transition. In order to establish a link between dynamics and structure,
it is worth analysing the evolution of some structural features as a function of time. For instance we can characterize the structural and dynamical heterogeneities discussed above 
by some corresponding correlation lengths and search for a connection between them.
To this end we estimate the correlation length associated to clusters of fast particles and to the icosahedral structures, respectively using the following relations: $\xi_{fast}\propto\langle N_c^{1/3}\rangle$ and $\xi_{ico}\propto\langle N_{ico}^{1/3}\rangle$. In addition we evaluate the spatial correlation length $\xi_6$ with fold-symmetry $l=6$ of the BOO, which can be extracted from the spatial correlation function

\begin{equation}
g_6(r)=\frac{4\pi}{13}\langle \sum_{m=-6}^{6}Q_{6m}(0)Q_{6m}(r)^* \rangle / \rho(r),
\label{eq:g6}
\end{equation}

\noindent via the  Ornstein-Zernike  expression $g_6\propto \frac{1}{r}\exp(-\frac{r}{\xi_6})$ . In Eq. \ref{eq:g6}, $\rho(r)$ is the radial density function.  The growing orientational correlation length can be characterized by a power-law function that diverges at the ideal glass transition $\phi_0$\cite{onuki2002phase}
\begin{equation}
\xi_6=\xi_0[(\phi_{0}-\phi)/\phi]^{-2/3}.
\label{eq:xi6}
\end{equation}
As suggested previously by Tanaka and coworkers \cite{tanaka2010critical} we can express the relaxation time $\tau_D$ in terms of the empirical Vogel-Fulcher-Tammann (VFT) expression, Eq.\ref{VFT} and $\xi_6$ can be fitted with Eq. \ref{eq:xi6}. Combining both, an analytic relation between $\tau_{D}$ and $\xi_6$ can be derived:

\begin{equation}
\log(\tau_{D}) \propto \xi_6^{3/2}.
\label{eq:tauvsxi}
\end{equation}
It is a reasonable assumption that also the other two structural correlation lengths can be described by a critical divergence analogue to Eq. \ref{eq:xi6} and hence we expect them to have a similar dependence on $\tau_{D}$.

\begin{figure}[h!]
\centering \includegraphics[width=0.6\linewidth]{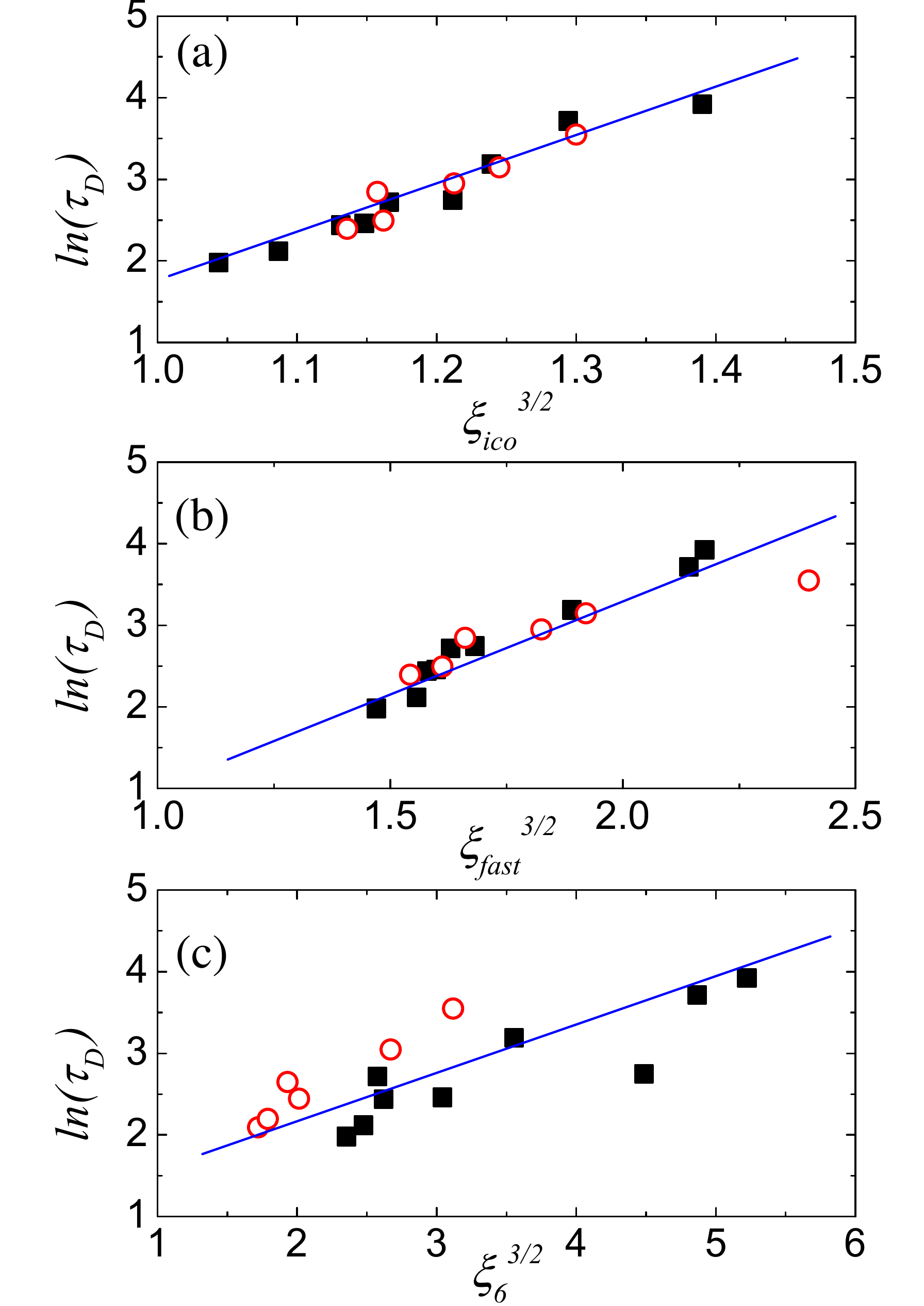}
\caption{Natural logarithm of the $\alpha$-relaxation time $\tau_{D}$ plotted versus three different spatial correlation lengths. Experiments: closed squares; simulations: open circles. (a) $\xi_{ico}^{3/2}$ - average size of icosahedral clusters $N_{ico}$; (b) $\xi_{fast}^{3/2}$ - average size of the fast particle clusters $N_c$ and (c) $\xi_6^{3/2}$ correlation length derived the bond orientational order parameter Q6. Lines are guides to the eye.}
\label{fig:tauvsxi}
\end{figure}

In Fig. \ref{fig:tauvsxi} (a)-(c) we verify the suggested scaling between $\tau_{D}$ and the three structural correlations both for
experiments and simulations.  For $\xi_{ico}^{3/2}$ and $\xi_{fast}^{3/2}$ a linear relationship with $\log(\tau_{D})$ is clearly confirmed. This shows that indeed the dynamics is strongly correlated with the appearance of icosahedral structures and clusters of fast particles. However, for $\xi_6$ such a connection is less evident. This in turn confirms that the crystalline bond orientational ordering does not play an important role in the dynamic slowing down of the system on approaching the glass transition.

Note that in our investigation the growth of the correlation lengths is found to be much smaller than the increase of the relaxation time. A recent work  on mixtures of hard-sphers has pointed out that the dynamic correlation length extracted from the overlap function  is always decoupled from the point-to-set correlation length, which represents an upper bound for the structural correlation lengths considered here\cite{charbonneau2013decorrelation}. Such results question the existence of a one-to-one causality relation between the growth of specific structures  and the dynamical slowing down  of the system close to the transition. Such correspondence has been also investigated with different tools coming from information theory \cite{jack2014information}  that confirm  a connection between LFS and mobility, although such correlation turned out to be weak. Hence, although our results suggest a link between dynamical slowing down and local structural correlations even in the absence of any crystal-like ordering,  the exact mechanisms connecting the growing static correlation lengths to the dynamic slowing down still remain a challenging question.
%Our results thus suggest a direct link between dynamical slowing down and local structural correlations even in the absence of any crystal-like ordering, differently from previous studies. However, the exact mechanism connecting the growing static correlation lengths to the dynamic slowing down still remains a challenging question.

\section{Summary and Conclusions}
In summary, we have presented a comprehensive study of the glass transition in emulsions that have moderate polydispersity. We have performed 3D confocal microscopy measurements over a range of volume fractions in order to sample the system below and well above $\phi_g$ up to jamming. The experimental study of a system in such an extended $\phi$ region, crossing the glass transition and even reaching marginal jamming conditions has been  previously attempted quite rarely. To obtain more detailed insights and to verify and benchmark our observations, we have compared our experimental results with a comprehensive set of Brownian dynamics simulations, finding remarkable agreement in all studied structural and dynamical properties. From this, we have demonstrated that uniform emulsions are excellent model systems for the study of the glass transition in soft colloidal systems.
\newline \indent In good agreement with previous work on hard spheres, we have observed that the dynamical slowing down on approaching $\phi_g$ is characterized by an increase of the relaxation time and the appearance of spatial and dynamical heterogeneities. The latter have been identified by the presence of fast and stable droplets that are spatially correlated. Fast droplets tend to form clusters whose size depend, not only on the distance from $\phi_g$, but also on time scale considered. A close link between the maximum cluster size and the relaxation time $\tau_{D}$ was observed. This suggests that fast droplets play an important role in the structural relaxation of the system. Analogously, mechanically stable droplets arrange in long-living clusters that have peculiar geometries. By performing topological cluster classification analysis we have shown that most of these clusters are icosahedra. Moreover, their population also increases on approaching the glass transition volume fraction, approximately saturating in the glassy region. The thorough investigation of these local and average properties at volume fractions below and above $\phi_g$ allowed us to follow the behaviour of structural and dynamical properties over a wide range, in- and out-of-equilibrium, finding that their all relevant parameters show a peak/dip or saturate at a maximum/minimum at $\phi_g$. We have also investigated whether the presence of a crystalline order exists and can be linked to the other structural signatures. Contrary to previous investigations on weakly polydisperse hard spheres, in our emulsion, which has a moderate polydispersity of about 12$\%$, the BOO parameters $Q_6$ and $Q_4$ do not increase either on approaching the transition or even above $\phi_g$; so,  we do not observe signatures for the onset of crystallization or of locally ordered crystal-like regions. Thus, we have been able to establish a clear link between growing structural correlation lengths and relaxation times, thereby confirming the existence of simultaneous structural and dynamical signatures of the glass transition even in the absence of the tendency to crystallize. Our results thus generalize the picture of heterogeneities occurring at the glass transition to the experimentally relevant case of polydisperse colloids and provide evidence that emulsions are a particularly advantageous model system for testing numerical and theoretical predictions.

%\emph{Acknowledgments}  
\ack
This research was supported by the Swiss National Science Foundation through project number 149867 (ZC, FS), by the European Research Council through project MIMIC, ERC Consolidator Grant number 681597 (NG, EZ), and by MIUR through Futuro in Ricerca project ANISOFT number RBFR125H0M (NG, EZ). TGM acknowledges support from UCLA.

\appendix
\section*{Appendix A}
The expression for the mean square displacement can be obtained assuming that the movement of particles in one direction is composed of two types of motions: rattling within the cage, and an inter-cage motion, such as hopping between different cages. Thus, the total displacement can be considered as the sum of
the inside cage term $c$ and an escape term related to the cage rearrangement  $h$: $\delta x=c+h$. We assume here that the two events are uncorrelated, therefore the 1D mean square
displacement is $\langle \delta x^2\rangle=\langle  c^2\rangle+\langle  h^2\rangle$.  It is reasonable to assume that the probability of finding a particle having distance $c$ from the center of the cage follows a Gaussian distribution $P(c) = N(0, \epsilon^2)$ with variance equal to the square of the cage size, from which follows that $\langle  c^2\rangle=\epsilon^2$. Analogously, the escape distance follows a Gaussian
distribution $P(h) = N(0, \zeta^2)$ where $\zeta$ is the characteristic cage-cage hopping size. Assuming that cage rearrangement is an
independent event then the  distribution of the number of cage rearrangement events $k$ follows the Poisson distribution: $P(k)=\lambda^k e^{-\lambda}/k !$ where $\lambda=t/\tau$ and $\tau$ is the lifetime of the cage. It follows that $\langle h^2\rangle=\sum_{k=1}^{\infty} \lambda^k e^{-\lambda}/k !  \,  h_k^2$, where $h_k$ is the expected displacement after $k$ cage rearrangements and, again, it follows a Gaussian distribution with zero mean and variance $k\zeta^2$. Then $\langle  h_k^2\rangle=k\zeta^2$, and $\langle  h^2\rangle=\lambda\zeta^2$. Following the previous considerations, we can write the 3D mean square displacement as $\langle \delta r^2\rangle=3\langle \delta x^2\rangle=3 [1+(\eta/\tau) t]\epsilon^2 = 3 (1+t/\tau_{D})\epsilon^2$ with $\eta=\zeta^2/\epsilon^2$. Note that we have defined the relaxation time of the system as $\tau_{D}=(\tau/\eta)$ and $3 \epsilon^2/\tau_{D}$ is by definition six times the diffusion coefficient of the system. By interpolating the mean square displacements with the expression derived above, we are able to extract $D$ and $\tau_{D}$ at several packing fractions $\phi$ . 
\newline

\bibliographystyle{unsrt}
\bibliography{EmulsionBib}

\end{document}